\documentclass[12pt,aps,prd,preprint,tightenlines,superscriptaddress,showpacs,nofootinbib]{revtex4-1}
\usepackage{graphicx}

\usepackage{amsmath} 
\usepackage{graphicx}
   \usepackage{amsmath}

\usepackage{subfigure}

\newcommand{\PRE}[1]{{#1}} 

\newcommand{\be}{\begin{equation}}
\newcommand{\ee}{\end{equation}}
\newcommand{\bea}{\begin{eqnarray}}
\newcommand{\eea}{\end{eqnarray}}

\newcommand{\lp}{\left(}
\newcommand{\rp}{\right)}

\newcommand{\nn}{\nonumber}

\newcommand{\ct}[1]{~\cite{#1}}

\newcommand{\tx}[1]{\mathrm{#1}}

\def\gev{\, {\rm GeV}}

\newcommand{\gsim}{\lower.7ex\hbox{$\;\stackrel{\textstyle>}{\sim}\;$}}
\newcommand{\lsim}{\lower.7ex\hbox{$\;\stackrel{\textstyle<}{\sim}\;$}}

\newcommand{\pb}{{\rm pb}}

\newcommand{\ig}[2]{ \raisebox{-0.5\height}{ \includegraphics[width=\textwidth,height=#1\textheight,keepaspectratio]{#2}}  }

\usepackage{enumerate}
\usepackage{amssymb,amsmath}

\begin{document}

\preprint{UH511-1235-2014}
\preprint{CETUP2013-021}

\title{
\PRE{\vspace*{1.3in}}
\textsc{MSSM dark matter and a light slepton sector: The Incredible Bulk}
\PRE{\vspace*{0.1in}}
}

\author{Keita Fukushima}
\affiliation{\mbox{Department of Physics \& Astronomy, University of
Hawai'i, Honolulu, HI 96822, USA}
}

\author{Chris Kelso}
\affiliation{\mbox{Department of Physics and Astronomy, University of Utah, Salt Lake City, UT  84112, USA}
\PRE{\vspace*{.1in}}
}

\author{Jason Kumar}
\affiliation{\mbox{Department of Physics \& Astronomy, University of
Hawai'i, Honolulu, HI 96822, USA}
}

\author{Pearl Sandick}
\affiliation{\mbox{Department of Physics and Astronomy, University of Utah, Salt Lake City, UT  84112, USA}
\PRE{\vspace*{.1in}}
}

\author{Takahiro Yamamoto \PRE{\vspace*{.1in}}}
\affiliation{\mbox{Department of Physics and Astronomy, University of Utah, Salt Lake City, UT  84112, USA}
\PRE{\vspace*{.1in}}
}

\begin{abstract}
Recent experimental results from the LHC have placed strong constraints on the masses of QCD-charged superpartners.
The MSSM parameter space is also constrained by the measurement of the Higgs boson mass, and the requirement that
the relic density of lightest neutralinos be consistent with observations.
Although large regions of the MSSM parameter space can be excluded by these combined bounds,
leptophilic versions of the MSSM can survive these constraints.
In this paper we consider a scenario in which the requirements of minimal flavor violation, vanishing $CP$-violation, and
mass universality are relaxed, specifically focusing on scenarios with light sleptons.
We find a large region of parameter space, analogous to the original bulk region, for which the lightest neutralino is a thermal relic
with an abundance consistent with that of dark matter.  We find that these leptophilic models are constrained by measurements of the magnetic and electric dipole moments
of the electron and muon, and that these models have interesting signatures at a variety of indirect detection experiments.

\end{abstract}

\maketitle

\section{Introduction}
Extensive searches for supersymmetric particles, including the lightest supersymmetric particle (LSP) dark matter candidate, have been carried out using a variety of approaches.
The absence of any direct production of sfermions at the Large Hadron Collider (LHC) excludes gluinos and 1st/2nd generation squarks
(if degenerate) with masses $\lesssim 1$ TeV,
while the sbottom and stop must be heavier than $\sim 100~\gev$~\cite{PDG}.
Moreover, the discovery of a Standard Model-like Higgs boson with mass $m_h \sim 126$ GeV \cite{Higgs} indicates
$m_{\tilde t} \gtrsim \mathcal{O}$(1 TeV)~\cite{Baer:2011ab}.
Within the constrained Minimal Supersymmetric Standard Model (CMSSM)~\cite{funnel,cmssm,efgosi,cmssm2,cmssmwmap}, the most well-studied, yet restrictive, supersymmetrization of the Standard Model,
one necessarily assumes universal boundary conditions for all scalar particles, gauginos, and trilinear scalar couplings at the scale at which the Standard Model gauge couplings unify.
This universality, along with the constraints
from data, together imply that all sparticles in the CMSSM must be relatively heavy~\cite{Kadastik:2011aa, Strege:2011pk,
Baer:2012uya, Akula:2012kk, Ghosh:2012dh, Fowlie:2012im, Buchmueller:2012hv, NUHM}.
If the LSP is a bino-like neutralino, as it is in much of the CMSSM parameter space, then large sfermion masses typically lead
to a small annihilation cross section, since processes mediated by sfermions are suppressed.  In the absence of coannihilation or some other
annihilation-enhancing mechanism, the resulting dark matter relic abundance would be far in excess of the value measured by the Planck satellite,
$\Omega_{\textrm{CDM}} h^2 = 0.1196 \pm{0.0031}$ \cite{Ade:2013zuv}.

In spite of these tight constraints on the squark masses and the very limited viable parameter space of the CMSSM, the current data still
leave open the possibility of models that possess relatively light sleptons and an electroweak-scale bino-like LSP, but with much heavier squarks.
Bounds on the masses of sleptons are much weaker than on squark masses; in particular,
the selectron, smuon, and stau need only be heavier than $\sim 100~\gev$~\cite{PDG}.
In this context, we examine a model that relaxes the standard assumptions of mass universality, minimal flavor violation, and $CP$-conservation for
the slepton sector.  Essentially, the parameters of $SU(3)_{\textrm{QCD}}$-charged sector of the theory and the leptonic sector will be decoupled; the $SU(3)$ sector will
be chosen to ensure consistency with collider searches and the Higgs mass measurement, while the leptonic sector will provide the annihilation channels
required in order for the lightest neutralino to have a thermal relic density that comports with astronomical observations.

However, for bino-like dark matter in the standard scenarios, there are well-known difficulties in obtaining an annihilation cross section large enough to sufficiently deplete the dark matter relic density.  In particular, the $s$-wave part of the annihilation cross section is
chirality-suppressed by a factor $\sim m_f^2 / m_\chi^2$, while the $p$-wave part of the cross section is velocity-suppressed by
a factor $v^2 \approx 0.1$ at freeze-out.  But if one departs from the assumption of minimal flavor violation in the slepton sector, then there can
be large mixing of left- and right-handed sleptons, which eliminates the chirality-suppression.  We then find a new
allowed ``bulk" region in a scenario that constitutes a minimal leptophilic version of the MSSM.

The great advantages of supersymmetric models include the elimination of quadratic divergences, precision grand unification, and the
presence of an acceptable dark matter candidate.  Frameworks such as the CMSSM also elegantly satisfy experimental constraints on flavor-changing
neutral currents (FCNCs).  The departure from the assumptions of the
CMSSM which we study will weaken some of these motivations; for example, because high-scale universality of soft supersymmetry-breaking mass parameters is relaxed, FCNC constraints are not automatically satisfied, and a little hierarchy
may be introduced.  However, given the
significant tension with data which exists in much of the parameter space of the CMSSM, it is quite possible that some of these motivations must
indeed be weakened in any model which can be consistent with observations.  The scenario we consider thus represents a relaxation of
the motivated assumptions underlying the CMSSM in a way which allows one to cleanly reconcile the model with data.
Supersymmetric constructions with spectra similar to those studied here have been shown to arise in supergravity grand unification models with gluino-driven radiative supersymmetry-breaking~\cite{Akula:2013ioa}.  
If one relaxes gaugino universality and takes the $\textrm{SU}(3)_C$ gaugino field to be much heavier than the other gaugino and sfermion fields at the unification scale, then the mass splitting between the electroweakinos and the gluino induces a mass splitting between the sleptons and squarks.  A similar mass spectrum can also be obtained in so-called split-family supersymmetry, where the sfermions of the first two generations are lighter than those of the third generation.  Explicit constructions leading to such spectra are discussed in~\cite{Ibe:2013oha} and~\cite{Evans:2013uza}.  Here, we explore the phenomenology of generic MSSM scenarios with electroweak scale bino-like dark matter and light sleptons.  This study also serves
as a guide for understanding the phenomenology of models with singlet fermion dark matter that couples to scalar leptons.

In this paper, we study the parameter space of this minimal leptophilic model, identify regions which are consistent with observational constraints, and
identify possible signals at current and upcoming experiments.
In Section II, we identify the key relationships between the parameters of the model (bino mass, slepton masses, mixing angles, and
$CP$-violating phases), and the relevant observables (the annihilation cross section and corrections to the $e/\mu$ mass and electric and magnetic
dipole moments).  In Section III we describe the current state of the relevant experimental bounds, arising from collider data,
dipole moment measurements, and indirect dark matter searches.  In section IV, we identify the regions of parameter space that
are consistent with current data, and prospects for finding evidence for such models at upcoming experiments.  We conclude with a discussion
of our results in Section V.

\section{Theory}

\subsection{Annihilation and the Relic Density }

In general, MSSM neutralino mass eigenstates are a mixture of bino and neutral wino and Higgsino states.
Neutral wino and Higgsino states couple directly to Standard Model gauge bosons, resulting in efficient annihilations, and therefore an
underabundance of dark matter, unless the pure wino or Higgsino LSP is relatively heavy.  Higgsino LSPs must be heavier than $\sim 1$ TeV
and wino LSPs must be heavier than $\sim 2.8$ TeV for the thermal relic abundance of each to explain the entirety of the
dark matter~\cite{Hisano:2006nn, Fan:2013faa}.  If dark matter is electroweak scale ($\sim100$ GeV) neutralinos, they must be dominantly bino-like.
In the absence of coannihilations, pure binos annihilate via sfermion exchange, which becomes less efficient as sfermions become heavier.
Electroweak scale bino-like dark matter is therefore a viable option so long as some sfermions are sufficiently light to mediate the annihilation.
This scenario with light sfermions and an electroweak scale bino-like neutralino LSP has long been termed the ``bulk'' region in the CMSSM~\cite{funnel},
though it has been excluded for some time because it predicts light squarks and a light $CP$-even Higgs boson with a low mass $\lesssim114$ GeV.

If scalar mass universality is assumed, as in the CMSSM, the absence of supersymmetric particles found at the LHC, the Higgs mass, and the
lack of experimental evidence of deviations from Standard Model expectations for rare decays all
force the sfermion masses to be heavy, as well \cite{Kadastik:2011aa, Strege:2011pk,
Baer:2012uya, Akula:2012kk, Ghosh:2012dh, Fowlie:2012im, Buchmueller:2012hv, NUHM}.
At the same time, the most natural version of neutralino dark matter, wherein no special mechanism such as coannihilation or resonance annihilation
is necessary to suppress the relic abundance into the cosmologically-viable range, would include light sparticles to mediate the annihilations
(see, eg.~\cite{King:2006tf}).  Indeed, within the CMSSM and other supersymmetric frameworks in which scalar mass universality is assumed, constraints
from colliders and cosmology are at odds.

The abundance of astrophysical cold dark matter is known to be
\bea
\Omega_{\tilde \chi}  h^2 = 0.1196 \pm{0.0031}
\eea
from the most recent measurements of the cosmic microwave background radiation~\cite{Ade:2013zuv}.
Assuming that dark matter is a thermal relic, the
freeze-out temperature is computed using the best fit solution to the Boltzmann equation, as in\ct{Bertone:2004pz},
\bea
x_f = \ln \left[ \sqrt{45 \over 8 } {m_{\tilde \chi} M_{\tx{Planck}}\langle \sigma v  \rangle_{x_f} \over g^{1/2}_*x_f^{1/2} \pi^3} \right],
\eea
where $m_{\tilde{\chi}}$ is the dark matter mass, $M_{\textrm{Planck}}$ is the Planck mass, $\langle \sigma v \rangle_{x_f}$
and $g_*$ are the thermally-averaged annihilation cross section and effective number of degrees of freedom, respectively, at freeze-out,
and $x_f = m_{\tilde{\chi}}/T_f$, where $T_f$ is the freeze-out temperature.
The cold dark matter relic abundance is then simply\ct{Gondolo:1990dk}
\begin{equation}
\Omega_{\tilde \chi} h^2 \simeq \frac{8.77 \times 10^{-11}~{\rm GeV}^{-2}}{\sqrt{g_*} \int^{T_f}_0 \frac{dT}{m_{\tilde{\chi}}} \, \langle \sigma v \rangle}.
\end{equation}
Since $x_f$ depends only logarithmically on the annihilation cross section,
the relic density is roughly inversely proportional to $\langle \sigma v \rangle_{x_f}$.  To obtain a relic density matching observation,
one would need $\langle \sigma v \rangle_{x_f} \sim 0.7~\pb$ (see also~\cite{Steigman:2012nb}), which is possible if sfermions are sufficiently light.
Relaxing scalar mass universality allows us to revive scenarios with electroweak-scale bino-like neutralino dark matter that annihilates via
light slepton exchange, while heavy squarks satisfy all collider constraints and boost the Higgs mass to the range measured at the LHC.
We refer to this scenario as the {\it new bulk region}.

\begin{figure}[t]
\center
\ig{0.15}{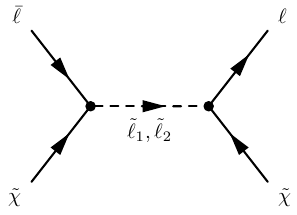} \ig{0.15}{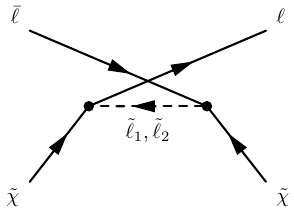}
\caption{The annihilation Feynman diagram}
\label{fig:annihilation}
\end{figure}

In the new bulk region, the leading dark matter annihilation channel is $\tilde{\chi} \tilde{\chi} \rightarrow \bar \ell \ell$, through
$t$-channel exchange of sleptons (this scenario is also considered in~\cite{Buckley:2013sca}, and
the $\ell = \tau$ scenario is discussed in detail in \cite{Pierce:2013rda}).
For simplicity, here we assume a pure bino LSP\footnote{Doping the LSP with some Higgsino content would enhance the annihilation cross section.
We do not consider this case further, except in a brief comment in Section~\ref{sec:analysis}.}.
The bino-lepton-slepton terms of the interaction Lagrangian are
\bea
L_{\tx{int}} =  \lambda_L {\tilde \ell}_L \bar{\tilde{\chi}} P_L \ell + \lambda_R {\tilde \ell}_R \bar{\tilde{\chi}} P_R \ell
+ \lambda_L^* {\tilde \ell}_L^* \bar \ell P_R \tilde{\chi}  + \lambda_R^* {\tilde \ell}_R^* \bar \ell P_L \tilde{\chi}  ,
\label{eq:Lint}
\eea
where the subscripts $L$ and $R$ denote the chiral eigenstates of the slepton. The slepton mass eigenstates are
related to the chiral eigenstates via the mixing parameter $\alpha$ by
\bea
\begin{bmatrix}  \tilde \ell_1 \\  \tilde \ell_2\end{bmatrix} =
\begin{bmatrix} \cos \alpha  & -\sin \alpha  \\ \sin \alpha  & \cos \alpha  \end{bmatrix}
\begin{bmatrix}   {\tilde \ell}_L \\  {\tilde \ell}_R \end{bmatrix}.
\eea
The $CP$-violating phase, $\varphi$, is absorbed in the coupling constants
\bea
\lambda_L  &=& \sqrt{2} g Y_L e^{ i  {\varphi \over 2} } ,
\nn \\
\lambda_R &=& \sqrt{2} g Y_Re^{-i  {\varphi \over 2} } ,
\eea
where the magnitudes of the constants are determined by the hypercharges $Y_L$, $Y_R$ and the hypercharge coupling $g$.
The Lagrangian in Eq.~\ref{eq:Lint} leads to the annihilation processes displayed in Fig.~\ref{fig:annihilation}.

Because bino annihilation exhibits no $s$-channel resonances,
one can expand  $\langle \sigma v \rangle$ in powers of $T/m_{\tilde{\chi}}$\ct{Srednicki:1988ce, Gondolo:1990dk} as
\begin{equation}
\left \langle \sigma v \right \rangle \sim  c_0 + c_1 \left ( \frac{T}{m_{\tilde{\chi}}} \right ) ,
\end{equation}
where $c_0$ is the velocity-independent $s$-wave contribution,
\begin{equation}
c_0 = {m_{\tilde \chi}^2 \over 2 \pi } g^4  Y_L^2 Y_R^2 \cos^2 \alpha \sin ^2 \alpha \lp {1 \over m_{\tilde \ell_1}^2
+ m_{\tilde \chi}^2} -{1 \over m_{\tilde \ell_2}^2  + m_{\tilde \chi}^2} \rp ^2 ,
\end{equation}
and $c_1$ is the $v^2$-suppressed contribution\footnote{The $v^2$-suppressed terms arise from both the
$s$-wave and $p$-wave matrix elements, but the $s$-wave terms will vanish in the $\sin (2\alpha) \rightarrow 0$ limit.},
\bea
 c_1= { m_{\tilde \chi}^2 \over 2 \pi } g^4 &\Biggl(& {(Y_L^4 \cos ^4 \alpha +Y_R^4 \sin ^4 \alpha )(m_{\tilde \ell_1}^4
 + m_{\tilde \chi}^4)\over (m_{\tilde \ell_1}^2 + m_{\tilde \chi}^2)^4} + { (Y_L^4 \sin ^4 \alpha +Y_R^4 \cos ^4 \alpha ) (m_{\tilde \ell_2}^4
 + m_{\tilde \chi}^4)\over (m_{\tilde \ell_2}^2  + m_{\tilde \chi}^2)^4}
\nn \\  &&
+ { 2(Y_L^4 +Y_R^4  )\sin ^2 \alpha  \cos ^2 \alpha (m_{\tilde \ell_1}^2 m_{\tilde \ell_2}^2 + m_{\tilde \chi}^4)\over (m_{\tilde \ell_1}^2
+ m_{\tilde \chi}^2)^2 (m_{\tilde \ell_2}^2  + m_{\tilde \chi}^2)^2}
\nn \\  &&
+ { Y_L^2 Y_R^2  \sin ^2 \alpha  \cos ^2 \alpha(m_{\tilde \ell_1}^2 - m_{\tilde \ell_2}^2 )^2\over 2(m_{\tilde \ell_1}^2
+ m_{\tilde \chi}^2)^4 (m_{\tilde \ell_2}^2  + m_{\tilde \chi}^2)^4}
 \Bigl[ 3 m_{\tilde \ell_1}^4 m_{\tilde \ell_2}^4 - 52 m_{\tilde \chi}^4 m_{\tilde \ell_1}^2 m_{\tilde \ell_2}^2 + 3 m_{\tilde \chi}^8
\nn \\  &&
- 14m_{\tilde \chi}^2(m_{\tilde \ell_1}^2 + m_{\tilde \ell_2}^2)( m_{\tilde \chi}^4+ m_{\tilde \ell_1}^2 m_{\tilde \ell_2}^2)
-5m_{\tilde \chi}^4(m_{\tilde \ell_1}^4 + m_{\tilde \ell_2}^4)  \Bigr]
\Biggr) .
\eea
Here we have assumed the fermion masses to be small, i.e.~$m_{\ell}/m_{\tilde \ell_i} \to 0$ (note that $c_0$ and $c_1$ do depend on $\varphi$
in terms proportional to $m_\ell$).
In the subsequent analysis, we will use the full $m_\ell$-dependent forms of $c_0$ and $c_1$.  The effect is only significant
for annihilations to $\tau$ leptons.

If $\sin (2\alpha )$ is small, then the $p$-wave term dominates the annihilation cross section, resulting in roughly
a factor of 10 suppression in the cross section at freeze out, and yielding a negligible annihilation cross section
in the current epoch.  But if $\sin (2\alpha) \sim {\cal O}(1)$, then the annihilation cross section can be unsuppressed
both at freeze-out and in the current epoch.

\subsection{Dipole moments}

\begin{figure}[t]
\center
\ig{0.2}{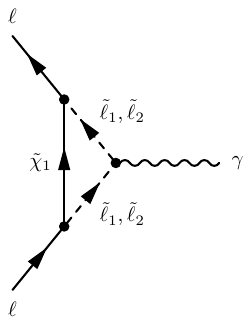}
\caption{The dipole moment Feynman diagram}
\label{fig:DM}
\end{figure}

In this scenario, a contribution to the electric or magnetic dipole moments of the Standard Model leptons can
arise from one-loop vertex correction diagrams with the bino and sleptons running in the loop, as shown in Fig.~\ref{fig:DM}.
Because the dipole moment operators flip the lepton helicity, the contributions to the dipole moments from
the bino-slepton loops can be large if L-R slepton mixing is large.
But the electric dipole moment can only receive a non-vanishing contribution if the $CP$-violating phase
$\varphi$ is non-zero.

In the limit $m_{\ell} \to 0$, the contributions to the anomalous magnetic moment, $a = \frac{g-2}{2}$, and the electric dipole moment,
$d/|e|$, of the associated lepton due to new physics are~\cite{Cheung:2009fc}
\bea
\Delta a &=&
 {m_{\ell} m_{\tilde \chi} \over 4 \pi^2 m_{\tilde \ell_1}^2} g^2 Y_L Y_R \cos \varphi
 \cos \alpha \sin \alpha   \left[ {1 \over 2(1- r_{1})^2} \lp 1+  r_{1} + {2 r_{1} \ln  r_{1} \over
 1- r_{1}}\rp   \right]
 -(\tilde \ell_1 \to \tilde \ell_2)
\nn \\
 {d \over |e|} &=& {m_{\tilde \chi} \over 8 \pi^2 m_{\tilde \ell_1}^2 } g^2 Y_L Y_R \sin \varphi
 \cos \alpha \sin \alpha  \left[ {1 \over 2(1- r_{1})^2} \lp 1+  r_{1} + {2 r_{1} \ln  r_{1} \over
 1- r_{1}}\rp   \right]
\nn \\  &&
-(\tilde \ell_1 \to \tilde \ell_2)
\label{eq:DMs}
\eea
where $ r_{i} \equiv m_{\tilde \chi}^2/m_{\tilde \ell_i}^2$.
Since charginos are assumed to be very heavy, diagrams with charginos and sneutrinos in the loop do not contribute.

Because dark matter annihilation from an $L=0$ initial state also requires a lepton helicity flip,
one may relate the $s$-wave part of the  $\tilde{\chi} \tilde{\chi} \rightarrow \bar \ell \ell$ annihilation cross section to
induced corrections to the electric and magnetic dipole moments of $\ell$.  In particular, in the
limit where the $r_{i}$ are small, we find
\bea
c_0 &\sim& 32 \pi^3 \left[ (\Delta a_\ell)^2 +\left({2m_\ell d_\ell \over |e|} \right)^2  \right] m_\ell^{-2}
+{\cal O}(r_{i})
\nonumber\\
&\sim& 3.9 \times 10^{11} \pb \left[ (\Delta a_\ell)^2 +\left({2m_\ell d_\ell \over |e|} \right)^2  \right] \left( m_\ell \over \gev \right)^{-2}.
\label{eq:dipole_cross_section}
\eea
This relation is similar to that found for scalar dark matter, as anticipated in~\cite{ADM}.
However, $c_0$ is maximized for $r_{i} \sim 1$, $r_{j \neq i} \rightarrow 0$.
In this limit, we find
\bea
c_0 &\sim& 72 \pi^3 \left[ (\Delta a_\ell)^2 +\left({2m_\ell d_\ell \over |e|} \right)^2  \right] m_\ell^{-2}
+{\cal O} \left( (1- r_{i})^2 \textrm{ or } r_{j} \right)
\nonumber\\
&\sim& 8.7 \times 10^{11} \pb \left[ (\Delta a_\ell)^2 +\left({2m_\ell d_\ell \over |e|} \right)^2  \right] \left( m_\ell \over \gev \right)^{-2}
\label{eq:dipole_cross_section}
\eea

\subsection{Mass correction }

\begin{figure}[t]
\center
\ig{0.13}{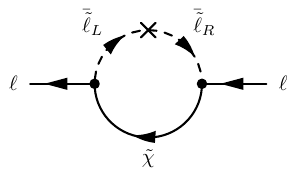}
\caption{The mass correction Feynman diagram}
\label{fig:mass_correction}
\end{figure}

If the slepton mixing angle $\alpha$
is allowed to be non-zero, then there will be a new contribution to the mass correction for Standard Model leptons arising from
the diagram in Figure~\ref{fig:mass_correction}, with the bino and sleptons running in the loop.
Note that this contribution includes a term which does not scale as the bare
lepton mass, implying that a small lepton mass is no longer technically natural.
Essentially, chiral symmetry no longer protects the fermion mass because it is broken by the scalar mass mixing term.
However, this correction is not logarithmic in the cutoff scale; the leading contributions from the
two diagrams with the two slepton mass eigenstates running in the loop cancel (analogous to the GIM mechanism).
This contribution to the mass correction is then given by
\bea
\delta m_\ell &\sim& {m_{\tilde \chi} \over 16 \pi^2 } Re (\lambda_L \lambda_R^*) \sin (2 \alpha )
\log \left({ m_{\tilde\ell_2} \over m_{\tilde\ell_1} } \right) ,
\eea
and leads to a little hierarchy problem with $\sim 1\%$ fine-tuning.
We make no attempt to address the flavor or naturalness problems of the Standard Model, however, and thus will not treat this
little hierarchy problem as an obstacle.
Similarly, we assume a priori that mixing is only allowed between sleptons of the same flavor, thus ensuring
that no new FCNCs are induced.

\section{Constraints}
As discussed previously, the Planck satellite has very accurately measured the dark matter
abundance to be $\Omega h^2 = 0.1196 \pm{0.0031}$ \cite{Ade:2013zuv}.  This requires the dark matter annihilation in any
particular scenario to be sufficient to ensure that dark matter was not over-produced in the early universe.  This
very restrictive constraint will limit the allowed parameter space in these models.

If the left-right slepton mixing angle is large, then one may potentially generate new charge-breaking
vacua.  If one demands that the charge-neutral vacuum be a global minimum, then very tight constraints
are imposed on the mixing angle.  But these conditions are weakened if one only requires that the
charge-neutral vacuum be metastable, with a lifetime as long as the age of the universe; even maximal mixing
is consistent with this metastability condition for the range of masses we consider~\cite{Hisano:2010re}.

We also examine several other
possible constraints as described below.

\subsection{Sparticle Mass Limits}
The current mass limits for supersymmetric particles are summarized by the Particle Data Group~\cite{PDG}.  The limits
(under some minimal assumptions) that the right handed scalar particles must satisfy are: selectrons must be heavier than 107\,GeV; smuons must be
heavier than 94\,GeV;  staus must be heavier than 82\,GeV; and sneutrinos must be heavier than 94\,GeV.
In addition, the LHC
has placed bounds on these particles through searches for their direct production.  Limits on the slepton mass (both left
and right handed selectrons and smuons) are presented by the ATLAS collaboration in Ref.~\cite{TheATLAScollaboration:2013hha}.  A left handed
slepton with a mass of 170 GeV to 300 GeV is excluded at 95\% confidence level for a neutralino mass of 100 GeV.  A right handed slepton
is still unconstrained for a neutralino mass of 100 GeV.  The CMS collaboration also performs a similar search with slightly weaker
results, although they only provide limits for the left-handed sleptons~\cite{CMS-PAS-SUS-13-006}.  As we are considering significant
mixing, these LHC limits can only be used as guidelines since a dedicated analysis including the effects of mixing would be required,
which is beyond the scope of this work.

\subsection{Indirect Detection}
The dark matter in our models will annihilate primarily to charged leptons (or neutrinos) leading to possibly detectable signals at
indirect detection experiments.  Strong constraints on dark matter annihilation in the current epoch
are placed on these models by gamma-ray searches as well as by the nonobservation of distinct bumps
in the otherwise rising positron fraction.

The Fermi collaboration has recently released the strongest constraints on dark matter annihilations
to leptons in the GeV to TeV mass range by looking at 25 Milky Way satellite galaxies~\cite{Ackermann:2013yva}.
There are currently no planned gamma-ray experiments that would lead to significant improvements on these indirect detection limits in the channels
relevant to our models. The Gamma-400 satellite will have significantly better angular and energy resolution than the Fermi-LAT, allowing
it to perform very sensitive searches for strong spectral features such as gamma-ray lines~\cite{Moiseev:2013vfa}.  The effective area
will be smaller, however, leading to only minor improvements in the limits in the channels relevant to our models~\cite{Bergstrom:2012vd}.

Strong constraints have also been derived from the nonobservation of bumps in the cosmic ray positron fraction due to dark matter
annihilations~\cite{Bergstrom:2013jra}.  Although the rise in the cosmic ray positron fraction remains unexplained, and could itself be due to
dark matter with a mass large enough that a cutoff in the spectrum is not yet observed, lighter dark matter ($\lesssim 100$ GeV) would produce a bump
with a cutoff at the dark matter mass if the annihilation rate is sufficiently high.  Because the data are of extremely high quality, and no such
bumps are observed, a limit on the annihilation cross section can be derived for any dark matter model.  For annihilations of 100 GeV WIMPs to charged
leptons, the constraints are near or even below the nominal thermal annihilation cross section, $(\sigma v)_{\textrm{th.}} \equiv 3 \times 10^{-26}$ cm$^3/$s.

We also consider light sneutrinos, which might possibly lead to a detectable annihilation signal in neutrino telescopes.  The
current IceCube limits are more than three orders of magnitude above the thermal relic scale for dark matter annihilations~\cite{Aartsen:2013mla}.
A preliminary result from Super Kamiokande has improved this limit significantly~\cite{SuperK}, but it still lies one to two orders of
magnitude above the thermal relic scale.

\subsection{CMB}
If dark matter particles annihilate at a sufficiently high rate, they could affect the CMB power spectrum.  One of the benefits of
this indirect detection technique is that it does not suffer from the astrophysical uncertainties of local signals.  The most recent
limits on the dark matter annihilation cross section for annihilations to several final states from the combination of various CMB, BAO, and supernovae
surveys are  given in Ref.~\cite{Madhavacheril:2013cna}.  For light WIMPs with masses $\lesssim 5$ GeV, the constraints are quite strong, disfavoring thermal annihilation cross sections.  However, the constraints weaken for larger WIMP masses.

\subsection{Dipole Moment Constraints}

In Table~\ref{table:dms}, we present the most recent measurements of the anomalous magnetic moments and electric dipole moments of the Standard Model
charged leptons~\cite{electron,Aoyama:2012wj,Karshenboim:2013opa,Baron:2013eja,muon,Bennett:2008dy,tau,Eidelman:2007sb,Abdallah:2003xd,Inami:2002ah},
along with the expectations for the anomalous magnetic moments within the Standard Model.  The leading order contributions to the electric dipole moments
of the charged leptons within the Standard Model occur only at more than three loops~\cite{Booth:1993af}, and are many orders of magnitude below the current
constraints.
\begin{center}
\begin{table}[h]
\caption{Measured Dipole Moments and Standard Model Expectations}
\begin{tabular}{| c | c | c |}
\hline
 & Measured Value & SM Expectation \\
\hline \hline
$a_e$ & $1159652180.76 (0.27) \times 10^{-12}$ & $1159652181.78(0.06)(0.04)(0.03)(0.77) \times 10^{-12}$ \\
$a_\mu$ & $116592091.(54)(33) \times 10^{-11}$ & $116591803(1)(42)(26) \times 10^{-11}$\\
$a_\tau$ & $117721(5) \times 10^{-8}$ & $-0.018(0.017)$\\
\hline
$\frac{d_e}{e}$ & $(-2.1 \pm 3.7 \pm 2.5) \times 10^{-29}  \,\tx{cm}$ &  \\
$\frac{d_\mu}{e}$ & $(-0.1 \pm 0.9) \times 10^{-19}  \,\tx{cm}$ &  $\sim 0$\\
Re$\left(\frac{d_\tau}{e}\right)$ & $(1.15 \pm 1.70) \times 10^{-17}  \,\tx{cm}$ & \\
\hline
\end{tabular} \label{table:dms}
\end{table}
\end{center}
The 2$\sigma$ ranges for the anomalous magnetic moments of the Standard Model charged leptons are
\begin{eqnarray}
\label{eq:mdmconstraints}
-2.66 \times 10^{-12} &<& \Delta a_e < 0.62 \times 10^{-12}
\nonumber \\
128 \times 10^{-11}  &<& \Delta a_\mu < 448 \times 10^{-11}
\\
-0.015  &<& \Delta a_\tau < 0.053,
\nonumber
\end{eqnarray}
while for the electric dipole moments, we have
\begin{eqnarray}
\label{eq:edmconstraints}
 -3.54 \times 10^{-18}  <  2 m_e\, {d_{e} \over e}  < 5.71 \times 10^{-18}
\nonumber \\
  -1.82 \times 10^{-6}  <  2 m_\mu\, {d_{\mu} \over e}  < 2.03 \times 10^{-6}
 \\
 -0.00405 < 2 m_\tau \,\textrm{Re}\left(\frac{d_\tau}{e}\right)  <  0.00819 .
\nonumber
\end{eqnarray}
Note that since we assume that the fundamental Lagrangian is $CPT$-invariant,
all dipole moments are real.  The precision of the measurement of the magnetic
dipole moment of the muon may be increased by up to a factor of 4 with data
from E821 at Fermilab~\cite{muon}.


\section{Analysis}
\label{sec:analysis}

In this leptophilic scenario, the squarks, charginos and neutralinos (except the bino) are all heavy and are chosen to
ensure that constraints from the Higgs mass measurement and direct sparticle searches at the LHC are satisfied.
This leaves
five relevant parameters to explore for a given slepton: $\alpha$, $\phi$, $m_{\tilde \ell_{1,2}}$, and $m_{\tilde{\chi}}$.
For each annihilation channel $\tilde{\chi} \tilde{\chi} \rightarrow \bar \ell \ell$, the annihilation cross section and dipole moment corrections
depend only on the superpartners of the particular $\ell$ (and their associated mixing angles and $CP$-violating phases), with no dependence on any
other sleptons.  As such, we can separately analyze each channel and simply sum the cross sections to determine the relic density if more than one channel is relevant.

\begin{figure}[h!]
\includegraphics[width=3.1in]{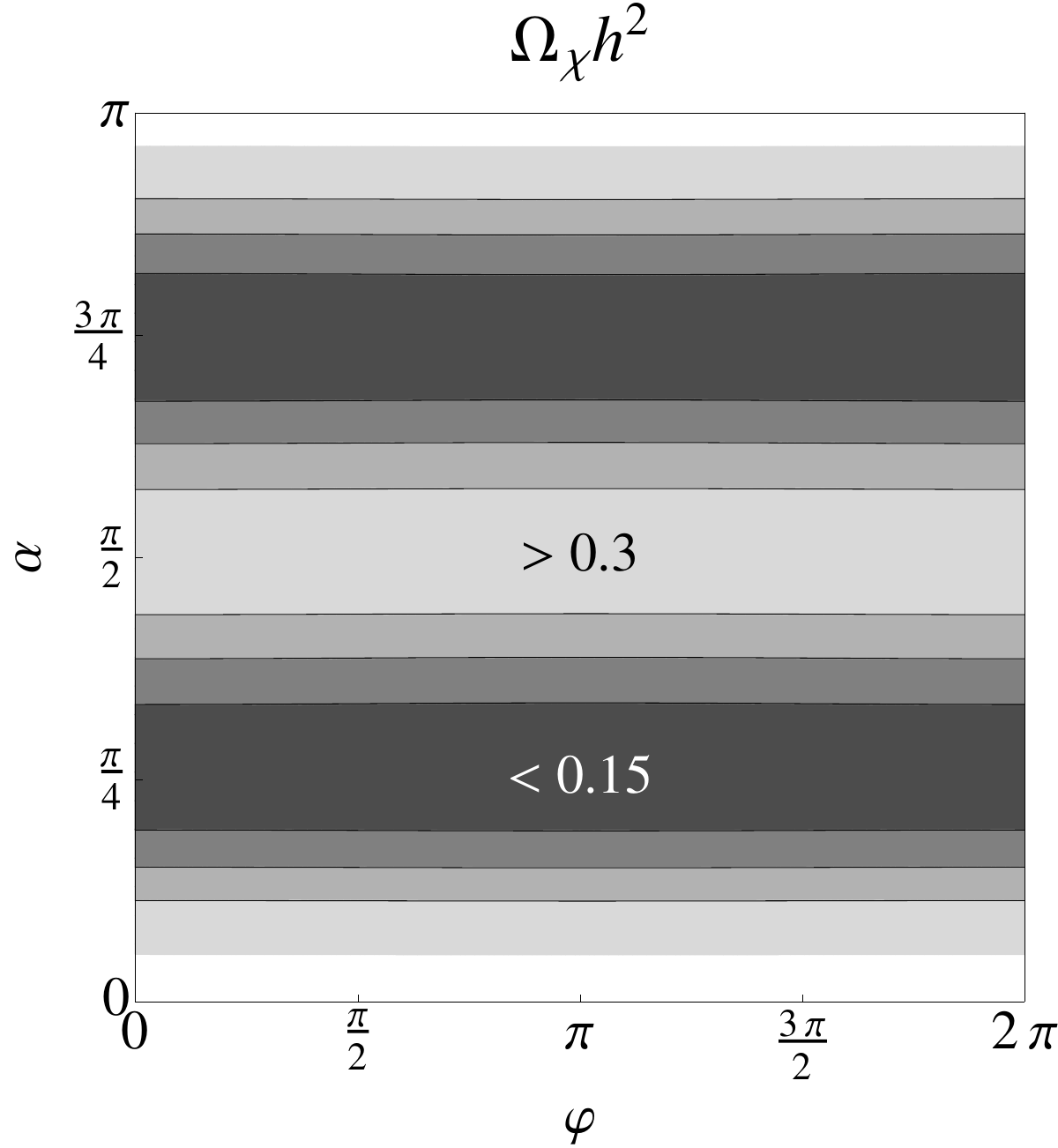}\hspace{4mm}
\includegraphics[width=3.1in]{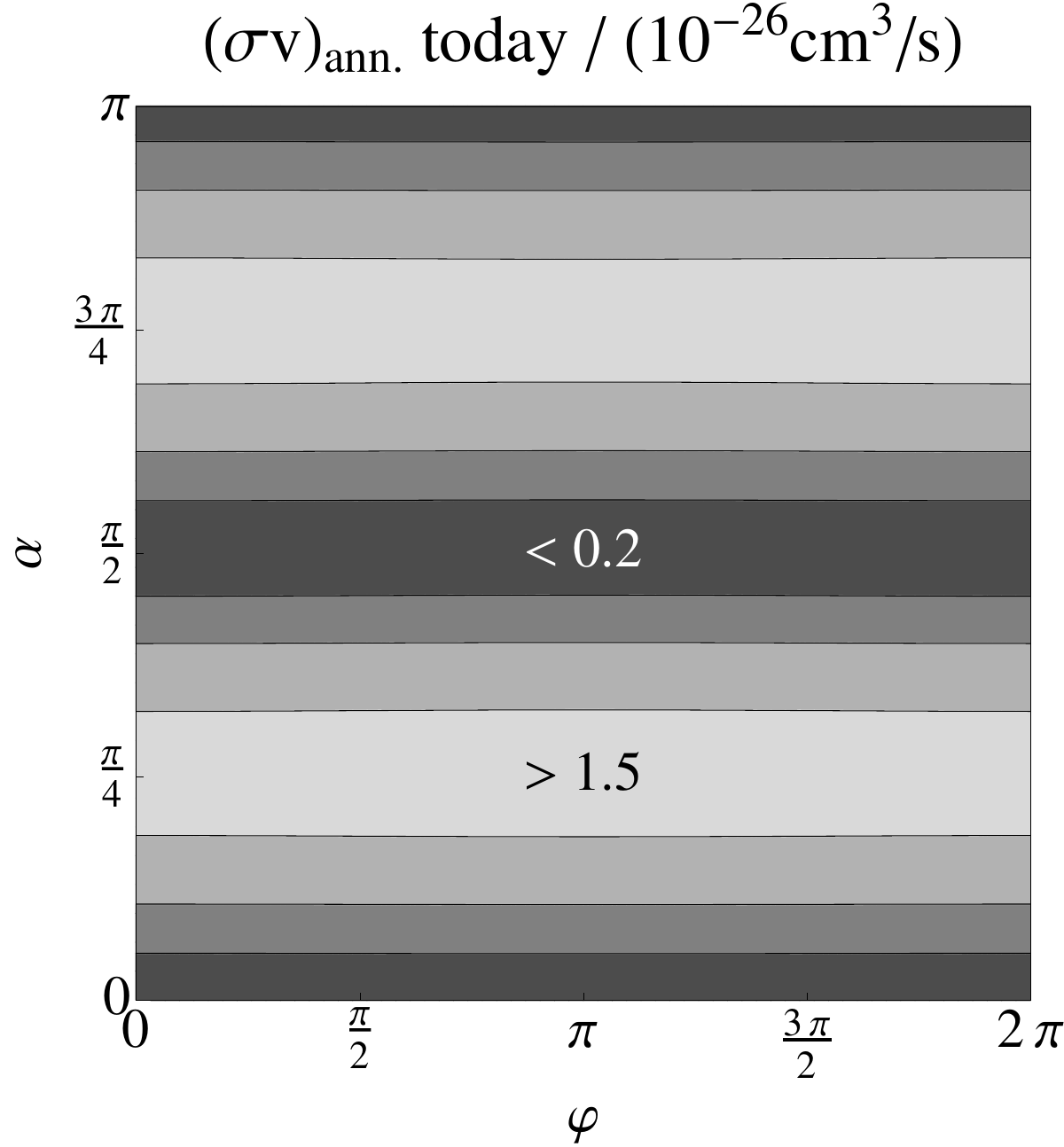}\vspace{4mm}
\includegraphics[width=3.1in]{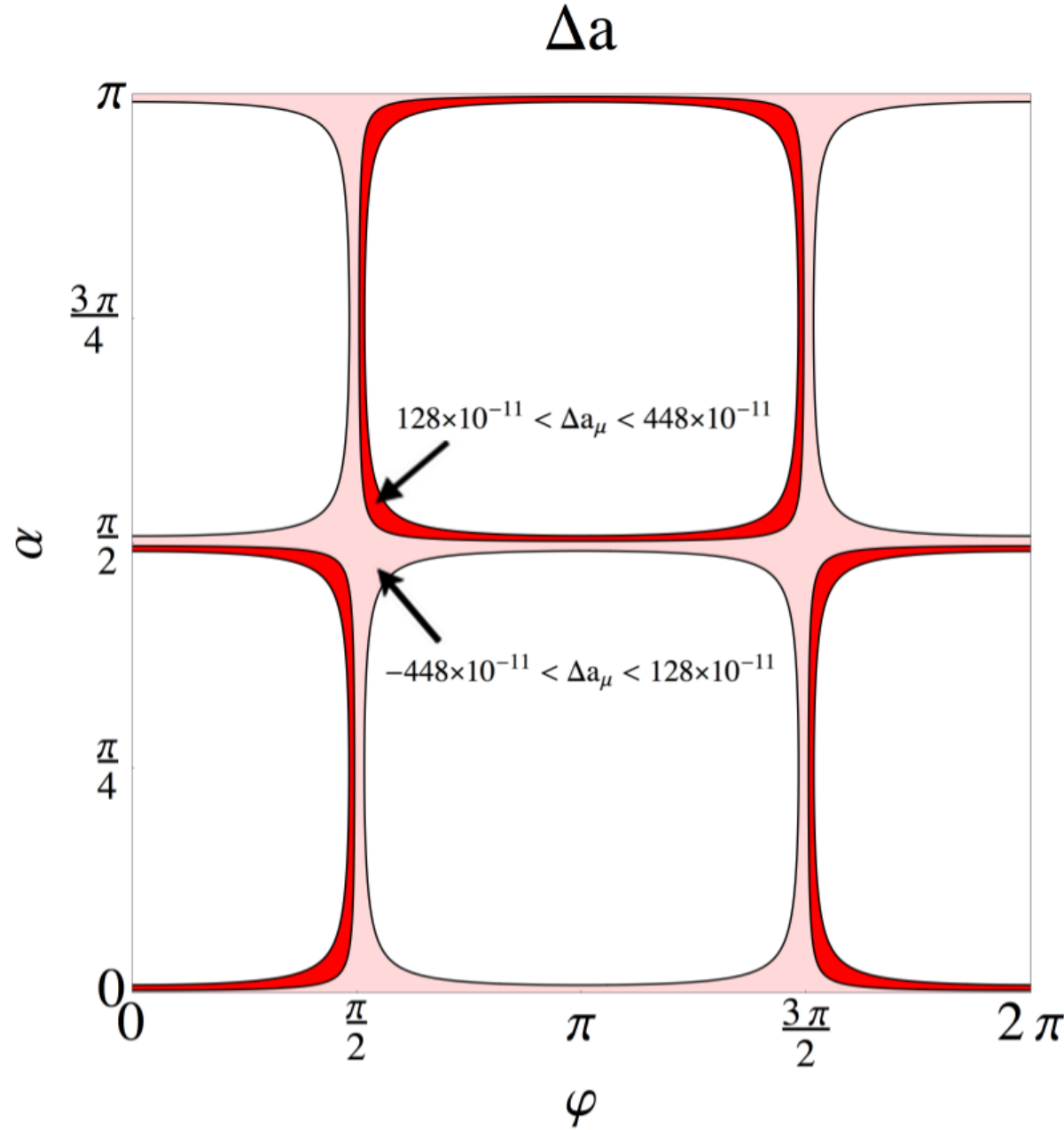}\hspace{4mm}
\includegraphics[width=3.1in]{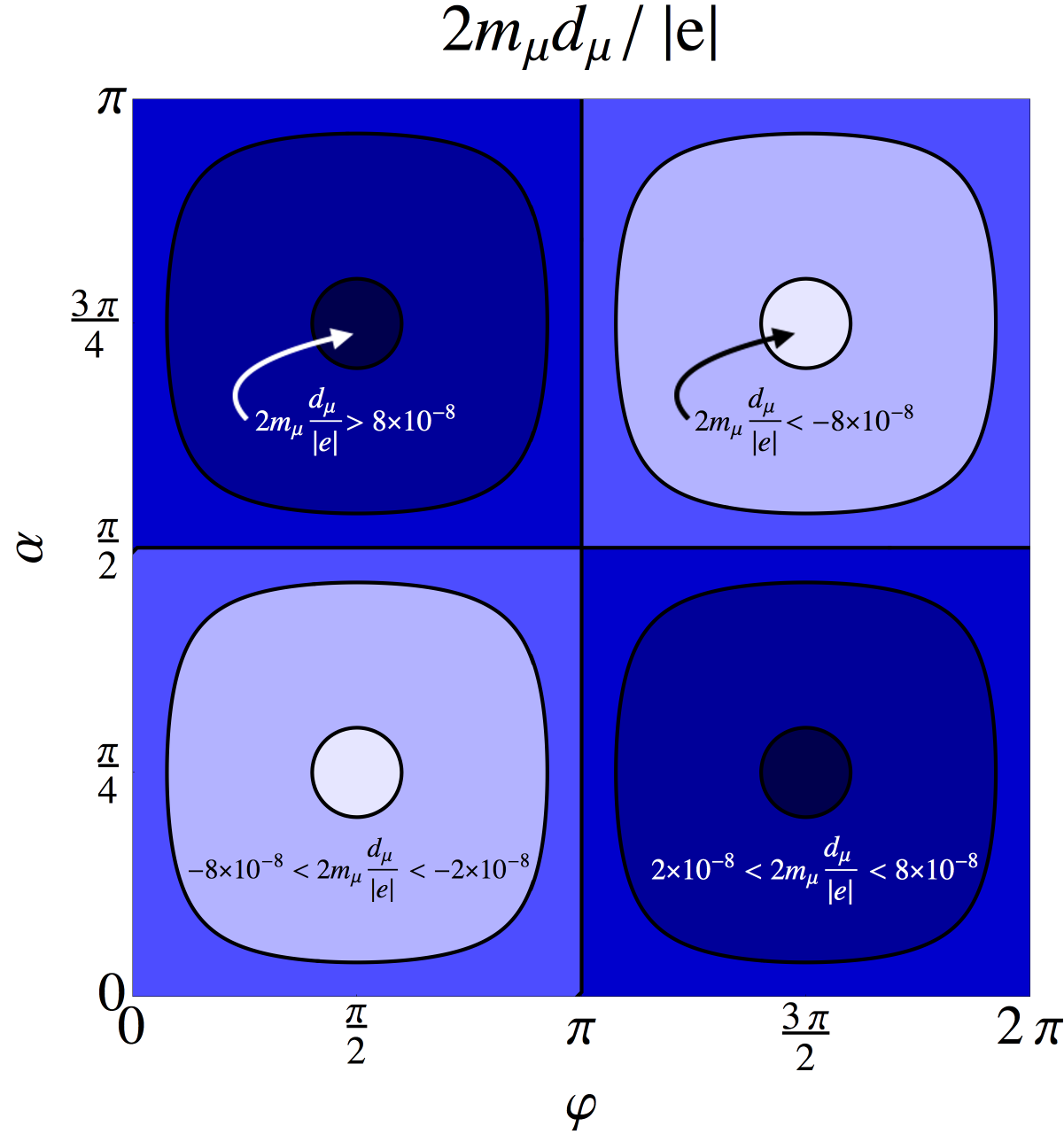}
 \caption{
 \label{fig:basicbehavior}
The dependence of various observables on the smuon L-R mixing angle, $\alpha$, and the CP-violating phase, $\varphi$, for
$m_{\tilde{\chi}} = 100$ GeV, $m_{\tilde{\mu}_1}=120$ GeV, and $m_{\tilde{\mu}_2} = 300$ GeV.
In the upper two panels, we present the neutralino relic abundance (left) and the neutralino annihilation cross section today (right).  In the lower panels, we present the
contribution to the anomalous magnetic (left) and electric (right) dipole moments of the muon.  In the darker red region in the lower left panel
this model fully accounts for the measured muon anomalous magnetic moment to $2\sigma$, while the lighter red shaded region provides a contribution
that is comparable to the measured value in magnitude.  In the lower right panel,
the electric dipole moment is unconstrained everywhere in the plane.}
\end{figure}

In Fig.~\ref{fig:basicbehavior} we show the dependence of various observables on $\alpha$ and $\varphi$ for a case with light smuons,
with all other sleptons heavy. In this example, $m_{\tilde{\chi}} = 100$ GeV, $m_{\tilde{\mu}_1} = 120$ GeV and $m_{\tilde{\mu}_2} = 300$ GeV.
In the upper two panels, we present the neutralino relic abundance (left) and the neutralino annihilation cross section today (right). In the lower panels of
Fig.~\ref{fig:basicbehavior}, we present the angular dependence of the new physics contributions to the
anomalous magnetic (left) and electric (right) dipole moments of the muon.

As expected, the annihilation cross section is maximized near maximal L-R squark mixing ($\alpha =\pi/4, 3\pi/4$) and is almost zero at $\alpha =0, \pi$,
while being nearly independent of $\varphi$.  The slight deviations from these above expectations arise from terms which scale as $m_\ell / m_{\tilde{\chi}}$;
these
terms are significant only for annihilation to the $\tau$ channel, and even then amount to roughly a $\sim5\%$ effect.
The dependence of the annihilation cross section on the slepton parameters is largely independent of the choice of
final state leptons; the distinction between annihilation channels arises instead from the experimental constraints on the slepton
masses and the dipole moment corrections.  In Fig.~\ref{fig:basicbehavior}, the thermal relic abundance is not quite low enough to be
within the $2\sigma$ Planck range for this particular choice of $m_{\tilde{\chi}}$ and $m_{\tilde{\mu}_{1,2}}$, though this could easily be
accomplished with only a small branching fraction to another final state, such as would occur if there were also a relatively light stau, or
with the addition of a very small Higgsino content to the LSP.  The annihilation cross section today clearly resembles the inverse of the relic
abundance, with expected values $\sim 10^{-26}$ cm$^3/$s.  Neither the relic abundance nor the annihilation cross section
today depend strongly on the fermion mass; therefore these results are approximately valid for all lepton final states.

In the lower panels of Fig.~\ref{fig:basicbehavior}, the planes are shaded according to the contribution to the anomalous magnetic moment of the muon (left) and
the electric dipole moment of the muon (right).  In the darker red region in the lower left panel this model fully accounts for the measured muon anomalous
magnetic moment to $2\sigma$, while in the lighter red shaded region new physics provides a contribution that is comparable to the measured value in magnitude.
The contribution
to the anomalous magnetic moment vanishes for $\varphi = n \pi/2$ for $n$ odd and for $\alpha =n\pi/2$ for integers $n$. In the lower right panel,
the electric dipole moment is unconstrained everywhere in the plane, and the shading indicates the value of the contribution to the muon electric dipole moment,
which reaches minima at $(\varphi ,\alpha)=(\pi/2,\pi/4)$ and $(3\pi/2,3\pi/4)$ of $2 m_\mu d/|e| \approx -10^{-7}$, and maxima at $(3\pi/2,\pi/4)$ and
$(\pi/2,3\pi/4)$ of $2 m_\mu d/|e| \approx 10^{-7}$.  These values are roughly an order of magnitude below the current sensitivity.
The contribution to the electric
dipole moment vanishes for $\varphi = n\pi$ and $\alpha = n\pi/2$ for integers $n$.  The only case in which contributions to both dipole moments vanish
is that of zero mixing.

\begin{figure}[t]
\includegraphics[]{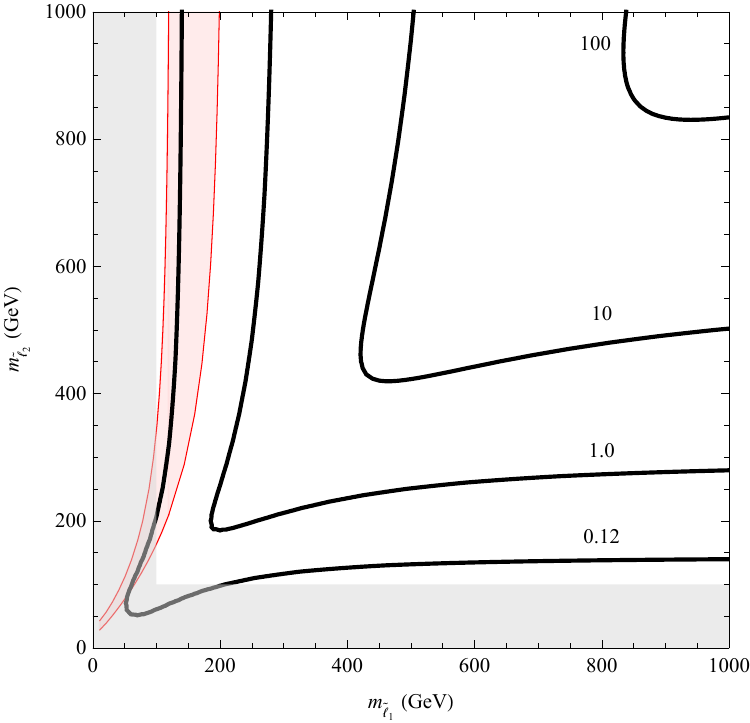}
 \caption{
 \label{fig:muonM1M2}
The dependence of the relic density and anomalous magnetic moment on the masses of the two smuons for the case of a 100 GeV
dark matter particle with $\alpha=\pi/4+0.02$ (the angle that minimizes the relic density) and $\varphi=\pi/2-0.04$.  The grey
region is disfavored because the smuon would be the LSP.  The labeled contours show the relic density and the red shading indicates
the $\pm 2\sigma$ region that would explain the measured value of $\Delta a_\mu$.  For this scenario, the size of the electric dipole
moment would be of order $10^{-9}$, which is well below the current limits and is left off the plot for clarity. }
\end{figure}

In Fig.~\ref{fig:muonM1M2}, we also examine how these observables depend on the masses of the smuons.  We again choose a dark matter
mass of 100 GeV and $\alpha=\pi/4+0.02$ (the angle that minimizes the relic density) and $\varphi=\pi/2-0.04$.  Within the grey regions,
one of the smuons would be the lightest supersymmetric particle (LSP).  The contours of constant relic density are shown
and the red shading indicates the $\pm 2\sigma$ region that would explain the measured value of the anomalous magnetic moment of the
muon.  For light smuons, the maximum size of the electric dipole moment would be of order $10^{-9}$, which is well below the current
limits and is not shown for clarity.  We are thus able to find viable regions of parameter space for the light smuons that satisfy all
constraints (including the relic density) and can even explain the measured value for the anomalous magnetic moment of the muon.

\begin{figure}[t]
\includegraphics[width=8cm]{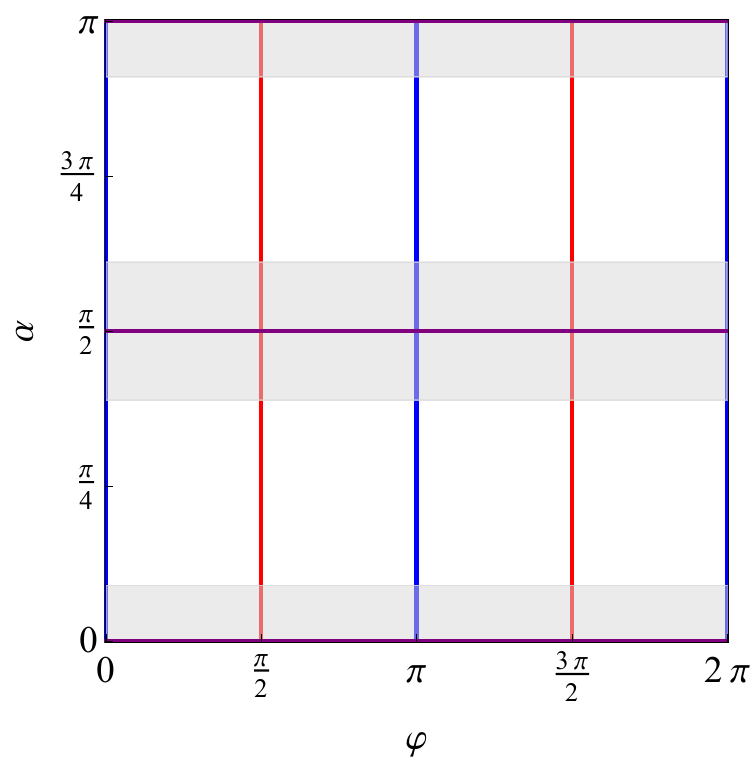}
\includegraphics[width=8cm]{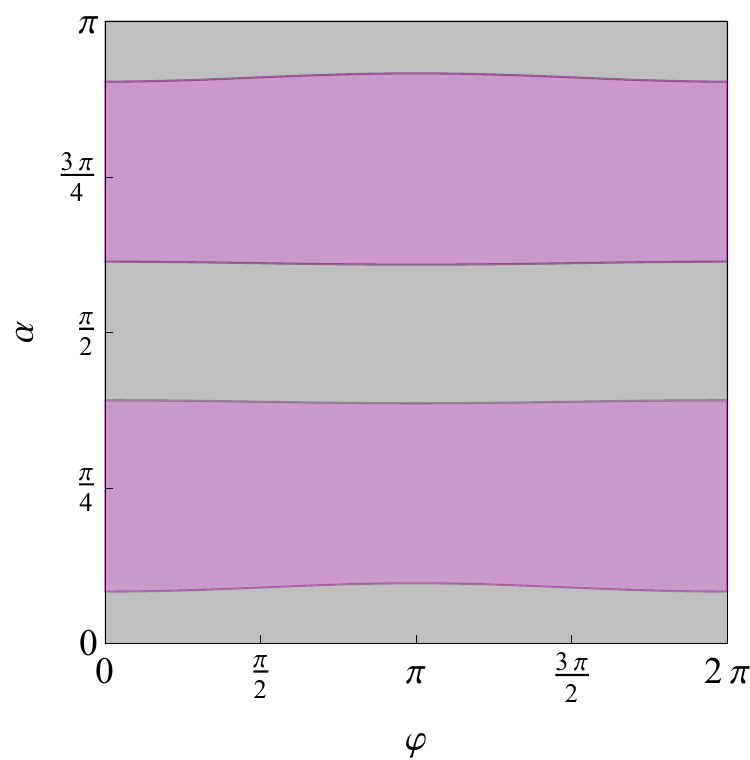}\vspace{5mm}
\includegraphics[width=9cm]{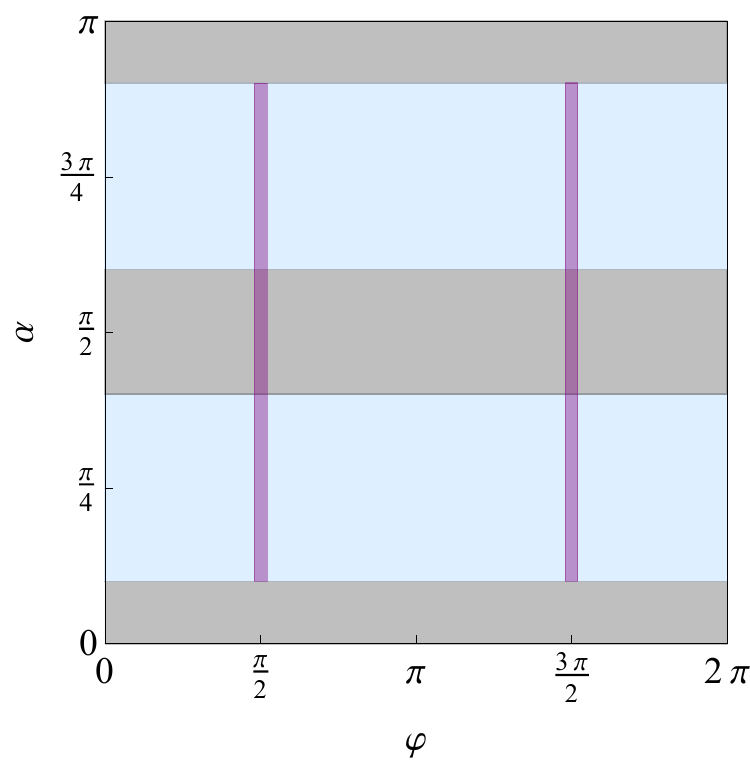}
 \caption{
 \label{fig:anglePlots}
The favored regions for the angles in the three different slepton cases
after marginalizing over the slepton masses.  The upper left is the
selectron, the upper right is the stau, and the bottom is for smuons.  In
each case, the greyed regions are excluded because the relic density would
exceed the $2\sigma$ measured value.  The blue (red) show the regions for
the angles where at least one $m_{\tilde \ell_1}, m_{\tilde \ell_2}$ mass
combination produces an electric (magnetic) dipole moment within the
current bounds.  For the selectron, there are no regions that satisfy all
three constraints, whereas for the stau, both
dipole moments constraints are too weak to provide any limits so the
regions are shaded purple (blue+red).  For the smuon case, the electric
dipole moment is not yet constraining these models, while the purple
shading shows $\left| \Delta a_{\mu} \right|\leq 4.5\times 10^{-9}$ (the
$\pm 2\sigma$ region that could explain the measured value
of the anomalous magnetic moment of the muon).}
\end{figure}

To gain insight into how general these features are, we can marginalize over the relevant masses and examine the constraints on the mixing angle
and $CP$-violating phase.  The three different charged lepton scenarios are shown in Fig.~\ref{fig:anglePlots}.
In each case, the greyed regions are excluded because the relic density would exceed the $2\sigma$ upper limit measured by Planck.  The blue
and red regions are favored by constraints on the electric and magnetic dipole moments, respectively.

Indeed, we see that the electron channel can be largely ignored for $s$-wave annihilation;
applying the constraints in eqns.~(\ref{eq:mdmconstraints}) and (\ref{eq:edmconstraints}) to
eqn.~(\ref{eq:dipole_cross_section}), we see that any choice of $\alpha$ and $\varphi$ yielding
$c_0 \sim 1~\pb$ would lead to a contribution to either the electric or magnetic dipole moment of the electron far in excess of
what could be consistent with experimental measurements, absent some large fine-tuning with other new physics contributions.
By contrast, annihilation to muons or taus can play a significant role in dark matter annihilation in the current epoch.
For muons, constraints on the anomalous magnetic moment force $\varphi \sim \pm \pi /2$.  In this case, the corrections
to the muon dipole moment are almost entirely $CP$-violating, and the contribution to the muon electric dipole moment is maximized
at $2m_\mu |d_\mu /e| \sim {\cal O}(10^{-9})$.  At this point in time, the muon electric dipole moment experiments
are much less constraining than those of the magnetic dipole moment, making this scenario perfectly viable.
For the $\tau$ channel, constraints on both the electric and magnetic dipole moments are too weak to be of any relevance.
As such, we see that the relic density constrains $\alpha$ while the dependence of the relevant parameter space on $\varphi$ is largely trivial.

In this analysis, we have explored only $m_{\tilde{\chi}}=100$ GeV.  If the neutralino mass is smaller,
then lighter sleptons must mediate the annihilations.  Scenarios with very light neutralinos with 1 GeV $<m_{\tilde{\chi}}<30$ GeV and
light sleptons have been explored in~\cite{Belanger:2012jn} and~\cite{Belanger:2013pna}, though both allow a non-negligible Higgsino content for the LSP.
They find neutralinos as light as 15 GeV with $\sim100$ GeV sleptons and $\sim200$ GeV charginos and next-lightest neutralinos are compatible with all current
collider constraints.  Since we focus here on $\sim100$ GeV dark matter, a more relevant question is how large the LSP mass could be while
remaining in the {\it new bulk}.  We find that if $m_{\tilde{\chi}} \sim m_{\tilde{\ell _1}}$,
then the mediating slepton must be lighter than $\sim150$ GeV,
making this a well-defined region with quite small charged sparticle masses\footnote{Note that this conclusion
is based specifically on the process $\tilde \chi \tilde \chi \rightarrow \bar \ell \ell$ via slepton exchange.  If
$m_{\tilde{\chi}} / m_{\tilde{\ell _1}} \approx 1$, coannihilation processes may be important, in which case heavier LSPs and heavier sleptons are possible.}.

Annihilation from an $s$-wave initial state can only be relevant to freeze-out (or current observations) if the
mixing terms are large.  Since right-handed neutrinos have no hypercharge, the only relevant $s$-wave annihilation
channels are those with two charged leptons in the final state.
The $s$-wave annihilation cross section can be expressed as
\bea
c_0
&\sim & (1.59~\pb)  \left({m_{\tilde \ell_1} \over 100~\gev} \right)^{-2} \sin^2 (2\alpha) \lp {2 \over r_{1}^{1/2} + r_{1}^{-1/2}}
- {m_{\tilde \ell_1} \over m_{\tilde \ell_2}} {2 \over r_{2}^{1/2} + r_{2}^{-1/2}  } \rp ^2
\eea
Assuming $m_{\tilde \ell_1} \leq m_{\tilde \ell_2}$, it is thus clear that $c_0$ is maximized
for $m_{\tilde \ell_2} \gg m_{\tilde \ell_1}$ and $\sin (2\alpha)=1$,
with $m_{\tilde \ell_1} \sim m_{\tilde{\chi}}$ and $m_{\tilde \ell_1}$ as light as possible, consistent with constraints from
data.
Interestingly, the maximum annihilation cross section is determined by the mass of the lightest slepton, and cannot
be increased by further decreasing the mass of the lightest neutralino; collider bounds on the slepton masses thus
place a firm bound on the $s$-wave annihilation cross section.
For staus or smuons with masses near 100 GeV and with maximal L-R mixing,
the $s$-wave annihilation cross section
may be ${\cal O}(1)~\pb$, providing a large enough annihilation rate to account for the dark matter relic density.

If $\sin (2\alpha) \ll 1$, then the $v^2$-suppressed terms in the matrix element become important.
In this limit,
\bea
\frac{c_1}{x_f}
&=&  ( 0.6~\pb )\left( {20 \over x_f} \right) \left({m_{\tilde \ell_1} \over 100~\gev} \right)^{-2}
\left( {8\left( Y_L^4 \cos ^4 \alpha +Y_R^4 \sin ^4 \alpha \right) \left( r_{1} + (1/r_{1}) \right)
\over \left(r_{1}^{1/2} + r_{1}^{-1/2} \right)^4} \right.
\nonumber\\
&\,& \left. + {r_{2} \over r_{1} }
{ 8\left( Y_L^4 \sin ^4 \alpha +Y_R^4 \cos ^4 \alpha \right) \left( r_{2} + (1/r_{2}) \right)
\over \left(r_{2}^{1/2} + r_{2}^{-1/2} \right)^4}
\right)
\nonumber\\
\eea
Assuming $m_{\tilde \ell_1} \leq m_{\tilde \ell_2}$, it is clear that these $v^2$-suppressed terms are maximized for $\sin \alpha=1$ in
the charged lepton annihilation channel, and for $\cos \alpha =1$ for the neutrino channel.  If
$m_{\tilde \ell_1} \sim m_{\tilde{\chi}}$, then annihilation to each charged lepton can provide a contribution
to the annihilation cross section of $\sim 0.6~\pb$; thus, annihilation to the $e$, $\mu$ and $\tau$
channels together can deplete the relic density enough to satisfy observational constraints.  Of course, if selectrons are light,
then the selectron $L-R$ mixing angle must vanish to one part in $10^{3}$ in order to satisfy the dipole moment constraints; that is,
if $\sin (2\alpha) \approx 0$, then the dipole corrections are small.  Note also that the contribution
of the neutrino channel is suppressed by a factor of 16 relative to the charged lepton channels, due to
the hypercharge of the left-handed neutrino.  Thus, sneutrino mediation can only provide a small contribution
to the total annihilation cross section for a viable model.
Although these $p$-wave annihilation channels can play a
significant role in dark matter annihilation at freeze-out, they have negligible impact on dark matter
annihilation in the present epoch.


The prospects for indirect detection of these models are quite good.
The sensitivity of the Fermi telescope to gamma rays from dwarf galaxies~\cite{Ackermann:2013yva} is relevant really only for the case of annihilations to taus:
a 100 GeV dark matter particle annihilating with the thermal cross section to $\bar \tau \tau$ would be roughly a factor of 5 above
the current Fermi limit. The
improved statistics due to the longer exposure and possible new dwarf galaxies discovered in the southern hemisphere by upcoming large
surveys makes this scenario potentially detectable.  The electron and muon cases are much less optimistic, however, with the current
limits
more than an order of magnitude above the thermal cross section.  Even the most optimistic assumptions would put these scenarios
just on the edge of detectability.

Strong constraints, however, have been derived from the AMS-02 measurement of the cosmic ray positron fraction~\cite{Bergstrom:2013jra}.
For annihilations of 100 GeV dark matter particles to $\bar \mu \mu$, $(\sigma v) \gtrsim 2\times 10^{-26}~{\rm cm}^3 /{\rm s}$
is excluded.  Though it is certainly
possible for our scenario to have escaped detection, a modest improvement in this constraint may completely exclude our case of light smuons, assuming
the only annihilation channel accessible is $\tilde{\chi} \tilde{\chi} \rightarrow \bar \mu \mu$.  For light staus, the current constraints are still a
factor of a few above the thermal annihilation cross section, so those models are viable and will remain so for quite some time.  And in all cases, the
constraints are uncertain by a factor of a few in either direction due primarily to the lack of knowledge about the local density and the energy losses
experienced by cosmic rays as they propagate throughout the Galaxy.
Of course, if annihilations proceed to more than one final state, for example with some nonzero branching fraction to both muons and taus, then the
constraints
from indirect detection weaken in proportion to the branching fraction.  We note that the constraints on lepton dipole moments would remain as
presented, as they
are not sensitive to the annihilation rate.

Finally, the most recent
limits on the dark matter annihilation cross section from the combination of various CMB, BAO, and supernovae
surveys are given in Ref.~\cite{Madhavacheril:2013cna}.  These constraints are
still roughly an order of magnitude above a detectable signal for thermal 100 GeV dark matter annihilating to charged leptons.
That analysis also shows that even a  cosmic variance limited CMB experiment would still be a factor of a few
above detection for a thermal 100 GeV dark matter particle.

\section{Conclusions}

We have considered a minimal
leptophilic version of the MSSM,
in which the parameters of the squark and slepton sector are decoupled.  In particular, it is
assumed that the squarks have large masses which are chosen to satisfy experimental constraints
from the Higgs mass measurement, direct collider searches and rare $B$ decays.  The
parameters of the bino-slepton sector can then be chosen to address the now decoupled
problem of achieving the correct dark matter relic density.  The relic density
depends only on the bino and slepton masses and the slepton mixing angle.  In this simplified
sector, there arises a new bulk-like region, in which the correct relic density is achieved
with bino dark matter that annihilates through light mediating sleptons.  The key region
of parameter space is $m_{\tilde{\chi}} \sim m_{\tilde \ell_1} \sim 100~\gev$, with maximal L-R squark
mixing.

The most relevant constraints on this scenario arise from direct slepton searches, and from
the contribution of bino-slepton loop diagrams to Standard Model lepton dipole moments.  In
particular, the only channels for which the dark matter annihilation cross section in the current epoch
could be $\sim 1~\pb$ (subject to the above constraints) are  $\bar \tau \tau$ and $\bar \mu \mu$.
Moreover, if $(\sigma v)_{\tilde{\chi} \tilde{\chi} \rightarrow \bar \mu \mu} \sim 1~\pb$, then there must be
large $CP$-violation in the smuon sector.  $p$-wave suppressed dark matter annihilation to electrons could
be relevant to dark matter freeze-out, but must be unobservably small in the current epoch.

It would be interesting to consider the implications for this scenario if AMS-02  were to find evidence for dark matter
annihilation to electrons or muons with $(\sigma v) \sim 1~\pb$,
but not to taus or hadronic states (which could be distinguished by the absence of the associated photons and/or antiprotons).  If interpreted within the framework of the MSSM, this data would imply
that the LSP was largely bino-like, since any significant Higgsino or wino fraction would result in the production
of hadronic final states which would yield anti-protons.  Moreover, if the only final states consistent with the
cosmic ray data were muons and electrons, then we could in fact conclude that the final state consisted of muons
and that the mass of the smuons must be relatively light (${\cal O}(100~\gev))$.  A large bino annihilation cross section
to electrons would imply light selectrons and L-R mixing, which is ruled out by the electron electric and magnetic dipole
moment bounds (absent some large fine-tuning).  Finally, we could conclude that there was large L-R smuon mixing (in
order to allow such a large annihilation cross section), and large $CP$-violation (in order to evade tight bounds
from the measurements of the muon magnetic dipole moment).

It is remarkable that so much information could be gleaned
about the parameters of the MSSM in this {\it new bulk} scenario with only data from AMS-02, even without new data from the LHC.
But this new bulk region could be sharply probed in the next physics run of
the LHC.  If the mass of the lightest slepton can be constrained to be larger than $\sim150$ GeV,
then it would not be possible to explain the observed dark matter relic density without
coannihilation in the early universe, and/or a non-trivial wino/Higgsino fraction.

\vskip .2in
{\bf Acknowledgments.}
We thank Kaustubh Agashe, Celine Boehm, Xerxes Tata, Luca Vecchi, and Christoph Weniger for useful discussions.
The work of JK is supported in part by DOE grant DE-SC0010504. CK and PS would like to thank Nordita for hosting the workshop ``What is Dark Matter'' where this work was partially completed.
JK would like to thank the University of Utah for its hospitality and partial support.
CK, JK, and PS would also like to thank CETUP* (Center for Theoretical Underground Physics and Related Areas), supported by the US Department of Energy under Grant No. DE-SC0010137 and by the US National Science Foundation under Grant No. PHY-1342611, for its hospitality and partial support during the 2013 Summer Program, where first discussions of this idea took place.

\bibliography{constraints}

\end{document}